\documentclass[twocolumn,showpacs,preprintnumbers,amsmath,amssymb,superscriptaddress]{revtex4}
\usepackage[toc,page]{appendix}
\usepackage{graphicx}
\usepackage{dcolumn}
\usepackage{bm}
\usepackage{color}
\usepackage{amsbsy}

\begin{document}

\title{Multiatom Greenberger-Horne-Zeilinger states using cavity-induced Rydberg blockade}

\author{Daniel Cano}
\affiliation{Instituto Madrile\~no de Estudios Avanzados, IMDEA Nanociencia, Campus Universitario de Cantoblanco, E-28049 Madrid, Spain} 

\author{J\'{o}zsef Fort\'{a}gh}
\affiliation{CQ Center for Collective Quantum Phenomena and their
Applications, Physikalisches Institut, Eberhard-Karls-Universit\"at
T\"ubingen, Auf der Morgenstelle 14, D-72076 T\"ubingen, Germany}

\author{Veronica Fernandez}
\affiliation{Spanish National Research Council (CSIC),
Institute of Physical and Information Technologies (ITEFI),
Serrano 144, E-28006 Madrid, Spain}

\author{Rodolfo Miranda}
\affiliation{Instituto Madrile\~no de Estudios Avanzados, IMDEA Nanociencia, Campus Universitario de Cantoblanco, E-28049 Madrid, Spain}
\affiliation{Departamento de F\'isica de la Materia Condensada, Universidad Aut\'onoma de Madrid, Cantoblanco, E-28049 Madrid, Spain}

\begin{abstract}
We describe a novel method for producing Greenberger-Horne-Zeilinger states in cold atoms coupled to a superconducting coplanar cavity. In the proposed scheme, atoms interact between each other by virtual photon exchange via a cavity mode. These interactions cause an asymmetric Rydberg blockade mechanism that suppresses simultaneous excitations into different atomic Rydberg levels, thus forcing all atoms to occupy the same Rydberg state. This mechanism has effect even if the atoms are separated by distances of the order of millimeters. The atomic populations are transferred adiabatically from the ground state into the entangled state by following a collective dark state with low dissipation rates.
\end{abstract}

\maketitle

\section{Introduction} \label{sec:introduction}

Entanglement is a fundamental tool for quantum information processing and quantum metrology. It can be realized in neutral atoms by means of the Rydberg blockade \cite{Jaksch:00,Saffman:10,Low:12}. The Rydberg blockade inhibits the excitation of more than one atom into Rydberg states whose energies are shifted by the interatomic interactions \cite{Singer:04,Tong:04}. The Rydberg blockade has been used in several experiments to generate atom-number entangled states \cite{Johnson:08,Urban:09,Gaetan:09,Wilk:10,Dudin:12,Ebert:14,Barredo:14} and to implement a controlled-NOT quantum gate \cite{Isenhower:10,Zhang:10}. These quantum operations were realized using a symmetric blockade mechanism, which considers the same Rydberg state for all the atoms.

Recent theoretical works \cite{Brion:07,Wu:10,Muller:09,Saffman:09,Carr:13,Cano:14,Liu:15} propose an asymmetric Rydberg blockade involving two different Rydberg states, $| r \rangle$ and $| v \rangle$, to perform more complex quantum operations. In the asymmetric Rydberg blockade, at least one of the three potentials $V_{rv}$, $V_{rr}$ or $V_{vv}$ has to exceed the excitation linewidth of the resonant laser field, while at least another of them has to be much smaller, depending on the desired quantum operation. Here, $V_{\mu \nu}$ denotes the potential between two atoms in the Rydberg states $| \mu \rangle$ and $| \nu \rangle$. These asymmetric potentials lead to the selective excitation of only those multiatom Rydberg states whose interatomic interactions are negligible compared to the excitation linewidth. The proposals include multiqubit quantum gates \cite{Brion:07,Wu:10}, mesoscopic gates \cite{Muller:09} and a variety of entangled states \cite{Saffman:09,Carr:13,Cano:14}, including the so-called Greenberger-Horne-Zeilinger (GHZ) state,
\begin{equation}
| GHZ \rangle = \frac{1}{\sqrt{2}} \left( |r,r,...,r \rangle + |v,v,...,v \rangle \right),
\label{Eq_GHZ}
\end{equation}
which is especially relevant due to its promising applications in atomic interferometers and quantum enhanced measurements \cite{Toth:14}. Unfortunately, the asymmetric Rydberg blockade finds fundamental obstacles as the atom number is increased, as has been demonstrated in a previous work \cite{Cano:14}. Increasing the atom number by increasing the atom density leads to level crossings between closely spaced atoms, which causes significant changes in $V_{rv}$, $V_{rr}$ and $V_{vv}$, thus diminish and finally destroying the entanglement process. On the other hand, increasing the atom number while keeping the interatomic separations results in a decrease of the interactions between distant atoms and the consequent reduction of the quantum operation speed.

The situation changes when the atoms are arranged in a cavity. Atoms located in the antinodes of the cavity mode, which mediates the interaction, do interact with each other independent of their spatial separation. Recent progress in loading cold atoms to the surface of a coplanar superconducting microwave cavity of centimeter size \cite{Bernon:13} paves the way to a new class of quantum systems with long range interactions. The electromagnetic field of the coplanar microwave cavity is confined to a small volume around its center conductor, which in combination with the strong electric polarizability of Rydberg atoms allows to reach strong coupling between atoms and a cavity mode. Theoretical proposals demonstrated entanglement between atoms and the cavity field \cite{Pritchard:14}, between atoms and solid state quantum circuits \cite{Xiang:13,Petrosyan:09}, and between distant Rydberg (super)atoms \cite{Petrosyan:08}, which can be robust against thermal photons in the cavity \cite{Sarkany:15}.

\begin{figure}
\centerline{\scalebox{0.25}{\includegraphics{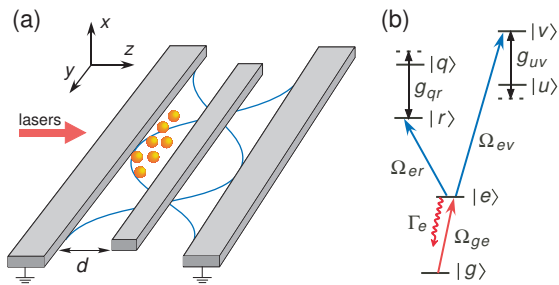}}}
\caption{(Color online) (a) A group of cold atoms are strongly coupled to a coplanar waveguide cavity. (b) Scheme of the atomic levels. Three laser fields are used to excite the atoms from the ground state $|g \rangle$ into two Rydberg states $|r \rangle$ and $|v \rangle$ using the fast-decaying, intermediate state $|e \rangle$. The Rabi frequencies $\Omega_{ge}$, $\Omega_{er}$ and $\Omega_{ev}$ correspond to the atomic transitions $|g \rangle \leftrightarrow |e \rangle$, $|e \rangle \leftrightarrow |r \rangle$ and $|e \rangle \leftrightarrow |v \rangle$ respectively. The population in $|e \rangle$ quickly decays to the ground state at a rate of $\Gamma_{e}$. States $|r \rangle$ and $|v \rangle$ are nonresonantly coupled to Rydberg levels $|u \rangle$ and $|q \rangle$ by means of the quantized field of the cavity. The atom-cavity coupling rates are $g_{pr}$ and $g_{qv}$ respectively.} \label{fig1}
\end{figure}

In this paper, we propose a novel class of collective coupling between cold atoms and a superconducting coplanar cavity. We demonstrate a cavity-induced asymmetric Rydberg blockade that allows to prepare the atomic ensemble in a GHZ state. Since the blockade is induced by the cavity, the entanglement of a large number of particles separated by macroscopic distances of millimeters or even centimeters can be achieved. The required laser intensities are compatible with the low-temperature conditions of superconducting devices. In this way, our method overcomes the problems associated with the asymmetric Rydberg blockade in free space \cite{Cano:14}. The proposed system is described in Fig. \ref{fig1}. It consists of $N$ neutral atoms trapped in the vicinity of a superconducting coplanar cavity operating in the microwave regime. The atoms are laser excited from the ground state $|g \rangle$ into two Rydberg levels $|r \rangle$ and $|v \rangle$ through the fast-decaying, intermediate state $|e \rangle$ by adiabatic rapid passage. In the proposed scheme, the cavity-mediated interactions are larger than the excitation linewidth if the atoms are in different Rydberg states, whereas they are smaller than the excitation linewidth if all atoms are in the same Rydberg state. In this way, the coherent laser excitation process leads the atoms into the entangled GHZ state.

\section{Cavity-induced asymmetric Rydberg blockade} \label{sec:potentials}

We study cavity-mediated interactions in an ensemble of $N$ cold atoms that are optically excited into two different Rydberg states, $| r \rangle$ and $| v \rangle$. To maximize the intensity of the interactions, we select $| r \rangle$ and $| v \rangle$ close to a F\"{o}rster resonance. This means that there are two other Rydberg states, $| q \rangle$ and $| u \rangle$, such that the atomic transitions $| r \rangle \leftrightarrow | q \rangle$ and $| v \rangle \leftrightarrow | u \rangle$ have similar frequencies, $\omega_{qr}$ and $\omega_{vu}$ respectively. These two atomic transitions are coupled nonresonantly to a cavity mode of frequency $\omega_c$. In the frame rotating with the cavity frequency, the Hamiltonian of the atom-cavity interaction is given by
\begin{eqnarray}
{\cal H}_{AC} &=& \hbar \sum_{j=1}^{N} \{ ( \Delta_v \hat{\sigma}_{uu}^{j} - \Delta_r \hat{\sigma}_{qq}^{j} )\nonumber \\
&+& ( g_{qr} \hat{\sigma}_{rq}^{j} \hat{a}_c + g_{uv} \hat{\sigma}_{vu}^{j} \hat{a}_c^{\dagger} + \text{H.c.} ) \} ,
\end{eqnarray}
where $ \hat{\sigma}_{\mu \nu }^{j} = | \mu_j \rangle  \langle \nu_j | $ is the transition operator of the $j^{th}$ atom, $\Delta_r=\omega_c-\omega_{qr}$ and $\Delta_v=\omega_c-\omega_{vu}$ are the cavity detunings from the atomic transitions, $\hat{a}_c^{\dagger}$ and $\hat{a}_c$ are the creation and annihilation operators of the cavity field, and $g_{\mu \nu} = - \wp_{\mu \nu} \varepsilon_c / \hbar$ are the atom-cavity coupling rates, which are calculated from the dipole matrix elements $\wp_{\mu \nu}=\langle \mu | \wp_ | \nu \rangle$ of the corresponding atomic transitions and the field per photon $\varepsilon_c = \sqrt{\hbar \omega_c / \epsilon_0 V_c }$ within the effective cavity volume $V_c=2 \pi d^2 L$. Here we have assumed that the coupling rates are the same for all atoms. At the atomic positions, between the central conductor and the ground plate, the electric field is parallel to the $z$-axis, so we use only the third component of the electric dipole operator, $\wp=ez$, where $e$ is the electron charge. This corresponds to $\pi$-transitions between atomic Rydberg states. To calculate the dipole matrix elements we use atomic wavefunctions obtained from a Numerov method \cite{Cano:12,Blatt:67,Bhatti:81} and the most accurately measured energies of the Rydberg levels \cite{Li:03,Mack:11}.

We consider the situation in which the cavity detunings, $\Delta_r$ and $\Delta_v$, are much larger than the F\"{o}rster defect, $\delta \omega = \omega_{qr}-\omega_{vu}$. This choice diminishes the probability of cavity field excitation by collective photon emission from atoms in state $| v \rangle$ \cite{Haakh:15}. In this way, the interaction between two atoms in their respective Rydberg states $| r \rangle$ and $| v \rangle$ is dominated by the van-der-Waals potential caused by virtual photon exchange. We simulate this situation with an ensemble of cold $^{87}$Rb atoms in realistic experimental conditions \cite{Bernon:13,Jessen:13,Megrant:12}. We select the Rydberg states $| r \rangle=| 67S_{1/2}, m_j=1/2 \rangle$, $| v \rangle=| 69S_{1/2}, m_j=1/2 \rangle$, $| q \rangle=| 67P_{1/2}, m_j=1/2 \rangle$ and $| u \rangle=| 68P_{3/2}, m_j=1/2 \rangle$. The frequency of the atomic transitions is of the order of $\omega_{qr} \simeq \omega_{vu} \simeq 2 \pi \times 11.9$ GHz, with a small F\"{o}rster defect of $\delta \omega = 2 \pi \times 2.8$ MHz \cite{Li:03,Mack:11}. These transition frequencies correspond to a cavity mode wavelength of $\lambda_c=\lambda \sqrt{\epsilon_r} \simeq$ 1cm, where $\lambda$ is the wavelength in vacuum and $\epsilon_r \sim$ 6 is the effective dielectric constant. Using the first longitudinal mode of the cavity, the length of the resonator is $L=\lambda_c$. For an electrode distance $d \simeq 10 \mu$m, the electric field per photon within the effective volume of the cavity is $\varepsilon_c \sim$ 0.37 V/m. With these parameters, the atom-cavity coupling rates are $g_{qr}=2 \pi \times 7.5$ MHz and $g_{uv}=2 \pi \times 10.4$ MHz. In a cavity with quality factor $Q\simeq 10^6$, the photon decay rate is $\kappa=\omega_c/Q \simeq$ 200 kHz \cite{Megrant:12}, and the condition for strong coupling is fulfilled. We also assume sufficiently low temperatures to neglect thermal photons in the cavity.

\begin{figure}
\centerline{\scalebox{0.3}{\includegraphics{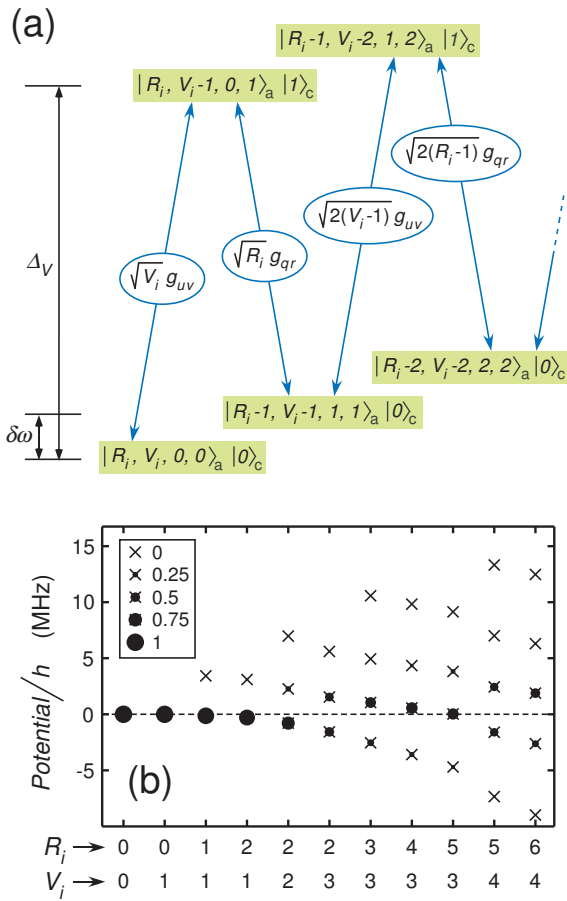}}}
\caption{(Color online) (a) Some of the collective Rydberg states $| R, V, Q, U \rangle_a | n \rangle_c$ involved in the atom-cavity interaction for a given initial state $| R_i, V_i, 0, 0 \rangle_a | 0 \rangle_c$. The coupling rates are written in the ovals. Although states with $n>1$ are not plotted, they are considered in the calculations. (b) Potential energy shifts of the zero-photon eigenstates $| R_i-\alpha, V_i-\alpha, \alpha, \alpha \rangle_a | 0 \rangle_c$, where $\alpha=0,1,...,min(R_i, V_i)$, for different values of $R_i$ and $V_i$. The graph represents an example of blocked excitation route in an ensemble of 10 atoms. The size of the circles represent $|C_i|^2$, where $C_i$ is the coefficient of the initial state $| R_i, V_i, 0, 0 \rangle_a | 0 \rangle_c$ in the respective eigenstate. The cavity detunings are $\Delta_r=2 \pi \times 110$ MHz and $\Delta_s=2 \pi \times 112.8$ MHz.} \label{fig2}
\end{figure}

We represent the Hamiltonian ${\cal H}_{AC}$ in the basis of symmetric states, $| R, V, Q, U \rangle_a | n \rangle_c$, where the capital letters $R$, $V$, $Q$ and $U$ denote the number of atoms in the respective Rydberg levels $| r \rangle$, $| v \rangle$ $| q \rangle$ and $| u \rangle$, and $n$ is the number of cavity photons. In this section we omit the number of atoms in $| g \rangle$ and $| e \rangle$ because these states are not coupled to the cavity field. Every initial configuration $| R_i, V_i, 0, 0 \rangle_a | 0 \rangle_c$  ($R_i + V_i \leq N$) is mixed with states $| R_i-\alpha, V_i-\beta, \alpha, \beta \rangle_a | \beta-\alpha \rangle_c$ ($\alpha=0,1,...,\beta$; $\beta=1,2,..., V_i$) by the Hamiltonian ${\cal H}_{AC}$. Figure \ref{fig2}(a) shows some of these states and the corresponding dipole elements. The dipole matrix elements in the symmetric-state representation are calculated as $\wp_{QR}=\langle R-1, V, Q+1, U | \wp | R, V, Q, U \rangle = \sqrt{(Q+1)R} \wp_{qr}$ and $\wp_{UV}=\langle R, V, Q, U | \wp | R, V+1, Q, U-1 \rangle = \sqrt{U(V+1)} \wp_{uv}$ \cite{Dicke:54}.

We calculate cavity-mediated interactions by diagonalizing ${\cal H}_{AC}$. Figure \ref{fig2}(b) describes the essentials of the cavity-induced asymmetric Rydberg blockade. It shows the potentials of a series of eigenstates of ${\cal H}_{AC}$ in an ensemble of ten atoms. Every column in the figure corresponds to eigenstates that mix with the same initial state, $| R_i, V_i, 0, 0 \rangle_a | 0 \rangle_c$, starting in the first column with all atoms in the ground state, $R_i = V_i = 0$, increasing the number of Rydberg excitations from left to right, up to the last column with all atoms in Rydberg states, $R_i = 6$ and $V_i = 4$. In this example, the excitation of all atoms into their corresponding Rydberg states is suppressed by the energy shifts in the sequence of intermediate states, even though some intermediate states are near resonance. Our simulations demonstrate that any excitation route with collective states containing atoms in both Rydberg levels, $| r \rangle$ and $| v \rangle$, is suppressed by the energy shifts caused by the interatomic potentials. Only excitation routes with the same Rydberg state for all atoms are in resonance with the excitation lasers.

The coupling of every eigenstate of ${\cal H}_{AC}$ to the classical laser fields is proportional to the projection of the eigenstate onto the corresponding initial state $| R_i, V_i, 0, 0 \rangle_a | 0 \rangle_c$. All the potentials in Fig. \ref{fig2}(b) have been corrected from the Stark shifts of the respective initial states $| R_i, V_i, 0, 0 \rangle_a | 0 \rangle_c$. These Stark shifts, which are calculated as $V_i |g_{uv}|^2 / \Delta_v$ from perturbation theory, are offset by detuning the classical field $\Omega_{ev}$, and have therefore no effect on the excitation process.

\section{Adiabatic transfer through collective dark states} \label{sec:dark_sates}

\begin{figure}
\centerline{\scalebox{0.3}{\includegraphics{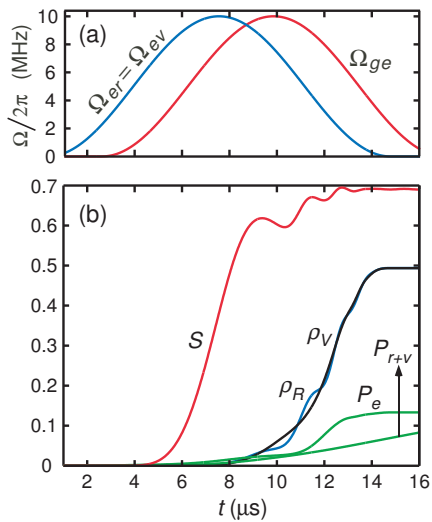}}}
\caption{(Color online) (a) Laser pulse sequence used for adiabatic transfer into Rydberg states. The temporal parameter is $\beta$ = 4.5 $\mu$s. (b) Numerical simulations of adiabatic transfer into the entangled state for $N=5$. The density matrix elements $\rho_R$ and $\rho_V$ represent the populations of the states $| R=N \rangle$ and $| V=N \rangle$ respectively, and $S$ is the von Neumann entropy of a single atom. The curves of $\rho_R(t)$, $\rho_V(t)$ and $S(t)$ consider only quantum trajectories that evolve without dissipation. $P_e$ and $P_{r+v}$ are the accumulated probabilities of spontaneous decay from the fast-decaying state $| e \rangle$ and from the Rydberg states respectively. The rate of spontaneous photon emission by the atoms sets the limit of the fidelity, which is $\cal F$ = 0.82 in the present example. The detunings are $\Delta_r = 2 \pi  \times 90$ MHz and $\Delta_v = 2 \pi  \times 92.8$ MHz.} \label{fig3}
\end{figure}

The atomic populations are transferred adiabatically from the ground states into the entangled states using adiabatic rapid passage \cite{Bergmann:98,Beterov:11}. The atom-field interaction Hamiltonian within the rotating-wave approximation is,
\begin{equation}
{\cal H}_{AL} = \frac{\hbar}{2} \sum_{i=1}^{N} ( \Omega_{ge} \sigma_{eg}^{i} + \Omega_{er} \sigma_{re}^{i} + \Omega_{ev} \sigma_{ve}^{i} + \text{H.c.} ),
\end{equation}
where the Rabi frequencies  $\Omega_{ge}$, $\Omega_{er}$ and $\Omega_{ev}$ have a typical temporal profile of adiabatic rapid passage,
\begin{equation}
\Omega_{\mu \nu}(t) = \left\{ \begin{array}{lr}
\Omega_{max} \cos^2\left(\frac{t-t_{\mu \nu}}{\beta}\right) \hspace{15pt} \text{if} \hspace{5pt} | \frac{t-t_{\mu \nu}}{\beta} | \leq  \frac{\pi}{2}
\hspace{2pt} , \\ 0 \hspace{15pt} \text{otherwise} , \end{array} \right.
\end{equation}
where the maximum peak amplitude $\Omega_{max}=2\pi \times 10$ MHz is the same along this work (see Fig. \ref{fig3}(a)). The temporal width $\beta$ is the same for the two laser pulses, although it may vary between different adiabatic transfer processes. In our model we choose $t_{er}=t_{ev}=t_{eg}-\beta/2$, so the relative delay between the two laser pulses equals $\beta/2$. This choice optimizes the adiabatic transfer efficiency into the Rydberg states.

To simulate the adiabatic transfer we transform the operators into the basis of eigenstates of $\cal H_{AC}$. In this way, the cavity-mediated interactions appear in the diagonal elements of the total Hamiltonian. Then, the dynamics of the system is simulated with the stochastic method of quantum trajectories \cite{Dalibard:92,Lambropoulos:06}. Each quantum trajectory represents the time evolution of the wave function, $|\Psi (t) \rangle$, in a particular thought experiment. We consider three decay channels: spontaneous decay from $| e \rangle$ into $| g \rangle$ at a rate of $\Gamma_{eg}=2\pi  \times 6.07$ MHz \cite{Petrosyan:13}, spontaneous decay from the Rydberg states at a rate of $\Gamma_{r} \simeq \Gamma_{v} \simeq 2\pi \times 0.4$ kHz \cite{Beterov:09} and the photon decay rate of the cavity $\kappa\simeq$ 200 kHz. During the simulation of each quantum trajectory, the computer randomly decides at every time whether the system spontaneously decays according to the current atomic populations and the cavity quantum field state.

\section{Multiatom entanglement in the microwave cavity} \label{sec:entanglement}

Figure \ref{fig3} shows the adiabatic transfer into the GHZ state for $N$=5. The populations of states $| R=N \rangle \equiv | r,r,...,r \rangle$ and $| V=N \rangle \equiv | v,v,...,v \rangle$, which are denoted by the respective density matrix elements $\rho_R(t)$ and $\rho_V(t)$, are represented in Fig. \ref{fig3}(b) for quantum trajectories that evolve without dissipation. We quantify the entanglement of the state using the von Neumann entropy $S(t)=-\langle\rho_1(t) \log \rho_1(t)\rangle$, where $\rho_1(t)$ is the reduced single particle density matrix. This is calculated as $\rho_1(t)=Tr_{(N-1)}\left[\rho(t)\right]$, where $Tr_{(N-1)}$ denotes the trace over the Hilbert spaces of $N-1$ atoms. The curve of $S(t)$ reveals a significant degree of entanglement already in the first moments of the process, when the atoms have not all been excited into the Rydberg states. At the end of the process the entropy reaches its maximum value log(2). The effect of dissipation is quantified by the accumulated probabilities of spontaneous photon emission from the intermediate state, $P_e$, and from the Rydberg states, $P_{r+v}$.

Our simulations show that the spontaneous emission of a single photon cancels the whole entanglement process. Indeed, the decay of a single atom from $|e\rangle$ projects the whole system into a collective state in which the other atoms are very likely to occupy their respective states $|e\rangle$. This is immediately followed by series of collective quantum jumps \cite{Lee:12}. Likewise, the collective spontaneous emission of a single photon from any of the Rydberg states acts as a measurement of the whole quantum system because all atoms tend to be in the same Rydberg state. The probability $P_e$ of spontaneous decay from $|e\rangle$ can be reduced by increasing the time parameter $\beta$, but this leads to higher probability $P_{r+v}$ of spontaneous decay from the Rydberg states. An alternative strategy would be to design the temporal profiles of the laser pulse to optimize the adiabatic transfer, but this task is beyond the scope of this paper. Our simulations also demonstrate that dissipation due to cavity photon loss is not relevant for the considered cavity detunings.

\begin{figure}
\centerline{\scalebox{0.3}{\includegraphics{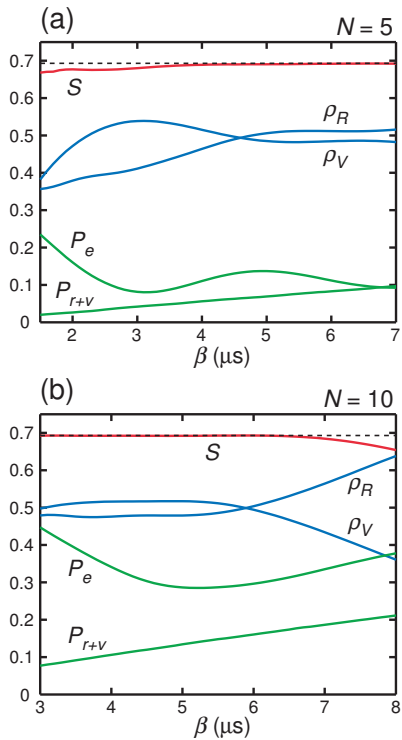}}}
\caption{(Color online) Populations $\rho_R$ and $\rho_V$, von Neumann entropy $S$, and accumulated decay probabilities $P_e$ and $P_{r+v}$ at the end of the laser pulse sequence for (a) $N=5$, $\Delta_r = 2 \pi  \times 90$ MHz and $\Delta_v = 2 \pi  \times 92.8$ MHz, and for (b) $N=10$, $\Delta_r=2 \pi \times 110$ MHz and $\Delta_s=2 \pi \times 112.8$ MHz.} \label{fig4}
\end{figure}

Figure \ref{fig4} summarizes the results for 5 and 10 atoms. It shows the populations $\rho_R$ and $\rho_V$, the von Neumann entropy $S$, and the accumulated decay probabilities $P_e$ and $P_{r+v}$ at the end of the laser pulse sequence as a function of the temporal parameter $\beta$. The curves of $\rho_R(t)$, $\rho_V(t)$ and $S$ consider only quantum trajectories without spontaneous emission. Increasing the atom number implies more decay channels that reduce the fidelity.

The final populations of $\rho_R$ and $\rho_V$ deviate slightly from the ideal value 1/2 as we vary $\beta$. This irregularity is caused by high-order interactions between atoms in the same Rydberg state $| v \rangle$. The importance of these interactions increase with the atom number. Another cause is the low interatomic potentials, which result in a weak blockade mechanism, in the collective states that contain only one atom in $| r \rangle$ ($| v \rangle$) and the rest of atoms in $| v \rangle$ ($| r \rangle$).

\section{Discussion} \label{sec:disscusion}

We have described an efficient method for producing multiatom GHZ states in atomic ensembles coupled to a superconducting coplanar cavity. The method relies on the Rydberg blockade induced by cavity-mediated interactions. The adiabatic transfer into the GHZ states is carried out through collective dark states. Although we have demonstrated our method using particular Rydberg states, it is not difficult to find such accidental F\"{o}rster resonances in the Rydberg spetra of alkalis, so our entanglement procedure can be generalized to other states.

Our proposal requires a good understanding and control over atom-surface interactions, especially the stray electric fields of adsorbed atoms and the Casimir-Polder potentials, in which significant progress has been made in the last years \cite{Tauschinsky:10,Abel:11,Hattermann:12,Crosse:00,Chan:14,Avigliano:14,Thiele:14,Jones:13,Sedlacek:16}. Additionally, stray potentials can be offset by detuning the classical fields $\Omega_{er}$ and $\Omega_{ev}$.

This work has been funded by a Marie Sklodowska-Curie Individual Fellowship (H2020-MSCA-IF-2014). Project reference: 660732. We thank Julio Camarero for his helpful comments.

\appendix
\section{Collective dark states}

To gain more insight into the adiabatic rapid transfer, we derive the equations of the collective dark states. The dark state is characterized by the destructive interferece between the quantum pathways from $|g \rangle$, $|r \rangle$ and $|v \rangle$ into $|e \rangle$. For a single atom this condition is expressed by the equation $\Omega_{ge} C_g+\Omega_{er} C_r+\Omega_{ev} C_v=0$, where $C_{\mu}$ are the coefficients of $|\mu \rangle$ in the atomic wavefunction. We now generalize this equation to ensembles of $N$ atoms. Since all atoms are initially in the ground state, we only consider dark eigenstates that are linear combinations of symmetric collective states $| G,E,R,V \rangle$, where $G$, $E$, $R$ and $V$ are the number of atoms in $| g \rangle$, $| e \rangle$, $| r \rangle$ and $| v \rangle$ respectively. In this appendix we omit the states that are not coupled by the lasers. The symmetric dark state is obtained from the set of equations $\sqrt{G} \Omega_{ge} C_{G,0,R,V}+\sqrt{R+1} \Omega_{er} C_{G-1,0,R+1,V}+\sqrt{V+1} \Omega_{ev} C_{G-1,0,R,V+1}=0$ for $G+R+V=N$, and $C_{G,E,R,V}=0$ for $E>0$, where $C_{G,E,R,V}$ denotes the coefficient of $| G,E,R,V \rangle$ in the wavefunction.

We now include the effect of the potentials by imposing the condition that $C_{G,E,R,V}=0$ if both $R$ and $V$ are higher than zero. Then, the equations of the dark eigenstates become
\begin{eqnarray}
\label{Eq_dark_states}
\sqrt{N} \Omega_{ge} C_{N,0,0,0} + \Omega_{er} C_{N-1,0,1,0} + \Omega_{ev} C_{N-1,0,0,1}=0,  \nonumber \\
\sqrt{G} \Omega_{ge} C_{G,0,R,0}+\sqrt{R+1} \Omega_{er} C_{G-1,0,R+1,0}=0 \nonumber \\
              \text{for     } G+R=N \text{     and     } R\geq 2, \\
\sqrt{G+1} \Omega_{ge} C_{G+1,0,0,V}+\sqrt{V+1} \Omega_{ev} C_{G,0,0,V+1}=0  \nonumber \\
              \text{for     } G+R=N \text{     and     } V\geq 2 . \nonumber
\end{eqnarray}

By elimination of terms from Eqs. \ref{Eq_dark_states} we find $(-1)^{N-1} \Omega_{ge}^N C_{N,0,0,0} + \Omega_{er}^N C_{0,0,N,0} + \Omega_{ev}^N C_{0,0,0,N}=0$. This expression connects the initial ground state, $C_{N,0,0,0}=1$, and the final GHZ state, $C_{0,0,N,0}=C_{0,0,0,N}=1/\sqrt{2}$.

\end{document}